\documentclass[12pt]{article}

\usepackage[english]{babel}
\usepackage{graphicx}
\usepackage{epsfig}
\usepackage{amssymb}

\newcommand{\beq}      {\begin{eqnarray}}
\newcommand{\eeq}      {\end{eqnarray}}

\newcommand{\bq}       {\begin{eqnarray}}
\newcommand{\eq}       {\end{eqnarray}}

\title
{The first $R_{cr}$ as a possible  measure of the  entrainment length in a 2D steady wake\\
}

\author{D. Tordella$^\sharp$ and S. Scarsoglio$^\sharp$}
\date{}

\begin{document}
\maketitle
{\it $^\sharp$  Dipartimento di Ingegneria Aeronautica e Spaziale,
Politecnico di Torino,  \textit{10129 Torino, Italy}}
\begin{abstract}
At a fixed distance from the body which creates the wake,
entrainment is only seen to increase with the Reynolds number ($R$) up
to  a distance of almost 20 body scales. This increase levels
up to a Reynolds number close to  the critical  value for the
onset of the first instability.  The entrainment is observed
to be almost extinguished at a distance which is nearly the same for all
the steady wakes within the  $R$  range here considered, i.e.
[20-100], which indicates that supercritical steady wakes have the same entrainment length as the subcritical ones. It is observed that this distance is equal to a number of body
lengths that is equal  to the value of the critical Reynolds number ($\sim 47$), as indicated by a large compilation  of experimental results. {\it A fortiori} of these findings,
we propose to interpret the unsteady
bifurcation as a process that allows a smooth increase-redistribution
of the entrainment along the wake according to the weight of the
convection  over the diffusion. The entrainment variation along
the steady wake has been determined using a matched asymptotic
expansion of the Navier-Stokes velocity field [Tordella and Belan,
{\it Physics of Fluids}, 15(2003)] built on
criteria that include the matching of the transversal velocity produced by the entrainment process.

\end{abstract}

\section{Introduction}

The dynamics of entrainment and mixing is of considerable interest
in engineering applications concerning pollutant dispersal or
combustion, but it is also of great relevance in geophysical and
atmospherical situations. In all these instances, flows tend to be
complex. In most cases, entrainment is a time dependent multistage
process in both  the laminar and turbulent regime of  motion.

The entrainment of  external fluid in a  shear flow such as that of  a wake or a jet is a
convective-diffusive process which is ubiquitous  when the
Reynolds number is greater than about a decade. It is a key
phenomenon associated to the lateral momentum transport in flows
which evolve about a main spatial direction. However, quantitative
data concerning the  spatial evolution of  entrainment are not
frequent in literature and are difficult to determine
experimentally. Quantitative experimental observations are very
cumbersome to obtain either in the laboratory or in the numerical
simulation context. In some cases, such as, for instance, fluid
entrainment by isolated vortex rings,  theoretical studies
(Maxworthy 1972\cite{Maxworthy1972}) predate experimental
observations (Baird, Wairegi and Loo 1977\cite{BairdAndWairegiAndLoo1977}; M\"{u}ller and Didden 1980\cite{MullerAndDidden1980}; Dabiri and Gharib 2004\cite{DabiriAndGharib2004}).

It is interesting to note that more attention has been paid to
complex unsteady and highly turbulent configurations in literature
than to  their fundamentally simpler steady counterparts.

In unsteady situations, entrainment is believed to consist of
repeated cycles of viscous diffusion  and circulatory transport.
In turbulent  flows,  a sequence of  processes is observed, where
the exterior fluid is first ingested by the highly stretched and
twisted interior turbulent motion (large-scale stirring) and is
then mixed  to the molecular level by the action of the
small-scale velocity fluctuations, see for instance the recent
experimental works carried out on free jets by Grinstein 2001\cite{Grinstein2001} or on a plane turbulent wake by Kopp, Giralt
and Keffer 2002\cite{KoppAndGiraltAndKeffer2002}.

In steady  laminar shear flows, stretching dynamics is
generally absent (as in 2D flows) or is close to its onset. In
this case, entrainment is determined by the balance between the
longitudinal and lateral nonlinear convective transport and the
mainly lateral molecular diffusion.

Air entrainment in free-surface flows is another important instance of the entrainment process. The mechanism is complex  and is also significant  in nominally steady flows ~e.g., a waterfall, or a steady jet. In such flows, entrainment is produced through the generation of cavities that can entrap air. The cavities are due to the impingement of the falling jet, which free-surface is usually strongly disturbed, over the liquid surface of the pool. As observed
in Ohl, Oguz, Prosperetti 2000\cite{OhlOguzPros00}, the generation process takes advantage of both the kinetic
energy of the jet surface disturbances and of part of the actual energy in the jet.


In this letter, we consider the steady two-dimensional (2D) wake flow past a circular
cylinder. We deduce the
entrainment as the longitudinal variation of the volume flow
defect using a matched Navier-Stokes asymptotic solution
determined in terms of inverse powers of the space variables
(Belan and Tordella 2002\cite{BelanAndTordella2002}; Tordella
and Belan 2003\cite{TordellaAndBelan2003}), see Section 2. This approximated (2D)
solution was obtained by recognizing the existence of a
longitudinal intermediate region,  which introduces the adoption
of the thin shear layer hypothesis and supports a differentiation
of the behaviour of the intermediate flow with respect to its
infinite asymptotics. The streamwise behaviour of the entrainment is presented in
Section 3.  The concluding remarks are given in Section 4.
\begin{figure}
  \includegraphics{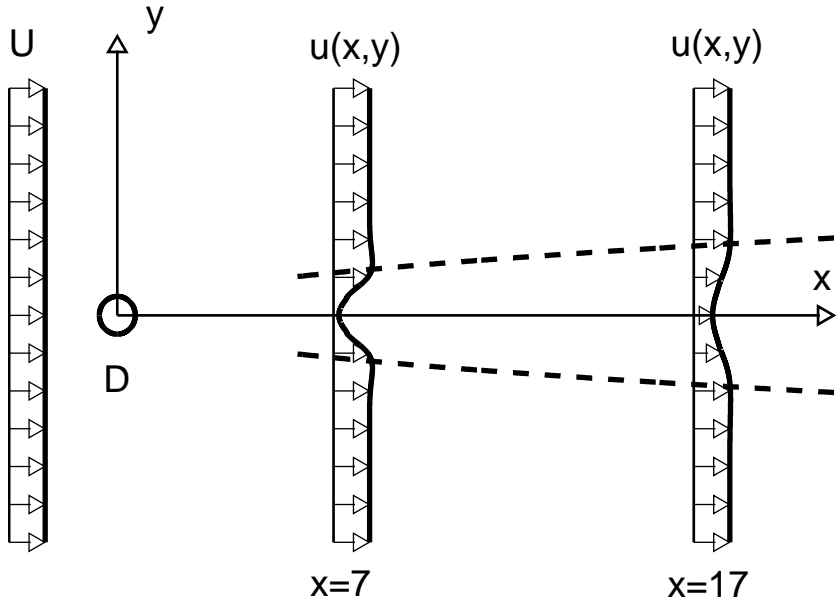}\\
  \caption{- Sketch of the physical problem. Longitudinal velocity
profiles (solid lines) at $R = 60$ and at stations $x = 7$, $x =
17$.}\label{sketch}
\end{figure}


\section{Analytical approximation of the velocity field, velocity flow rate defect and entrainment}
For an incompressible, viscous flow behind a bluff body, the
adimensional continuity and Navier-Stokes equations are expressed
as
\begin{eqnarray}
   u \partial_x u + v \partial_y u + \partial_x p &=& R^{-1} \nabla^2 u \\ \label{NS}
   u \partial_x v + v \partial_y v + \partial_y p &=& R^{-1} \nabla^2 v \\
   \partial_x u + \partial_y v &=& 0
\end{eqnarray}
where $(x, y)$ are the adimensional longitudinal and normal
coordinates, $(u, v)$ the adimensional components of the velocity
field, $p$ the pressure and $R$ the Reynolds number. The physical
quantities involved in the adimensionalization are the length $D$
of the body that generates the wake, the density $\rho$ and the
velocity $U$ of the free stream, see the flow schematic in fig. 1.
The Reynolds number is defined as $R = \rho U D/\mu$, where $\mu$
is the dynamic viscosity of the fluid.

The velocity field for the intermediate region of the
$2D$ steady wake behind a circular cylinder was obtained by matching an inner solution - a
Navier-Stokes expansion in negative powers of the inverse of the
longitudinal coordinate $x$
\beq f_i = & f_{i 0}(\eta) + x^{-1/2} f_{i 1}(\eta) + x^{-1} f_{i
2}(\eta) + \cdots \label{ei} \eeq
\noindent where $f$ is a generic dependent variable and where the quasi-similar transformation $\eta = x^{-1/2}
y$ is  introduced, and an outer solution, which is a
Navier-Stokes asymptotic expansion in powers of the inverse of the
distance $r$ from the body
\beq f_o = f_{o 0}(s) + r^{-1/2} f_{o 1}(s) + r^{-1} f_{o 2}(s) +
\cdots \label{eo} \eeq
\noindent where $r = \sqrt{x^2 + y^2}$ and $s = y/x$.

The wake mass-flow deficit of the inner field was considered by means of an {\it infield} boundary condition carefully accounting for it. In fact, this condition is placed at the beginning of the intermediate flow region which  inherits the full dynamics properties of near field. To this aim, we took advantage of  experimental velocity and pressure profiles, as usually done in many physical contexts and as suggested, in the present context,  by Stewartson (1957)\cite{Stewartson1957}. Further details about the use of this infield condition are given below. It should be noted  that the matched expansion in ranging from minus infinity to plus infinity in the transversal flow direction and that the concept of wake flow is clearly defined downstream from the intermediate region where the thin layer hypothesis starts to apply.
The relevant boundary conditions involve, aside the infield condition,  symmetry to the longitudinal coordinate and uniformity at infinity, both laterally and longitudinally.  For details on the expansion term determination, the reader can refer to  Tordella
and Belan 2003\cite{TordellaAndBelan2003}.

The physical quantities involved in the matching criteria are the
vorticity, the longitudinal pressure gradients generated by the
flow and the transversal velocity produced by the mass entrainment
process. The composite expansion is defined as $f_{c} = f_{i} +
f_{o} - f_{common}$, where $f_{common}$ is the common part of the
inner and outer expansions.  In Tordella
and Belan 2003\cite{TordellaAndBelan2003} the explicit inner and outer
velocity and pressure  expansions can be found up to order
four (i.e. $O(x^{-2})$ and $O(r^{-2})$, for the inner and outer
wake, respectively), the composite approximation has been
shown graphically. In this work, we approximate the wake
flow with the composite  solution obtained by
truncating the inner and outer expansions at the third order term
and then by determining their  common part by taking the inner
limit of the outer approximation. For the reader's convenience, the
inner and outer velocity component expansion terms  are listed below (see equations 9 - 23). The common part has not been included because it has an
analytical representation which alone would take up  a few pages.
However, the {\copyright \it Mathematica} file that describes its analytical
structure and which allows its computation is given in the EPAPS online
repository\cite{FileCommon}. The common expansion was obtained by writing the inner and outer
expansions in the primitive independent variables and by taking
the inner limit of the outer expansion, that is, by taking the
limit for $s \rightarrow 0$ and $r \rightarrow \infty$. To this
end, the  Laurent series of the outer expansion about $x
\rightarrow \infty$ was considered up to  the first order.
The composite expansion - which is, by construction, a
continuous curve, since it is obtained  by the {\it additive
composition} of three continuous curves,  the sum of the inner and
outer expansions  minus the part they have in common - is accurate
if the common expansion is accurate. This is always  obtained if,
at each order, the distance $\delta_n = | f_{i,n} - f_{o,n} |$
between the inner and the outer expansions is bounded and is at most of the same order as the range of $f_i$ and $f_o$. In the
present matching, we  have verified that in the matching region -
that is, in the region where the composite connects the inner and
the outer expansions - the distance $\delta_n$ is not only
bounded, but  is small with respect to the ranges of $f_i$ and
$f_o$.




The velocity  approximation is shown in figures 2 and 3, where the longitudinal and
transversal components of the composite solution for the velocity
field are plotted for different
longitudinal stations and Reynolds numbers.

It should be noted, that in this analytical flow representation, a few key properties of the wake flow 
have been taken into account.  These properties  can help an accurate description of the entrainment process to be obtained. These properties are:

i) The existence of intermediate asymptotics for the wake flow, in the general sense as given by Barenblatt and Zeldovich\cite{BarenblattZeldovich1972}. 
This is an important point, because the existence of  the
intermediate region  supports the adoption of the thin shear layer
hypothesis and relevant near-similar variable transformations
for the inner flow, while, at the same time, it also supports a
differentiation of the  behavior of the intermediate flow with
respect to its infinite asymptotics (Oseen's flow).

ii) The  use of an in-field boundary condition which
consists of the distribution of the momentum and pressure at a given
section along the mainstream  of the flow in opposition to the use
of integral field quantities.  This kind of boundary condition is
not new in literature\cite{Stewartson1957}, and presents the evident advantage of having a higher
degree of field information with respect to the use of integral
quantities such as the drag  or the lift coefficients (a given
integral value can be obtained from many different distributions).

iii) The acknowledgment of the fact that in free flows, such as
low Reynolds numbers - 2D or axis-symmetric - wakes or jets
developing in an otherwise homogeneous and infinite expanse of a
fluid, the main role in  shaping the flow is played by the inner
flow. This directly inherits the main portion of the  convective
and diffusive transport of the vorticity, which is created, at the
solid boundaries, by the motion of the fluid relative to the body.
For these flows, it is physically opportune to denote 
the "inner" flow as the straightforward or basic
approximation. This means that, up to the first order, the inner
solution is independent of the outer solution. According to this,
the Navier-Stokes model, coupled with the thin layer hypothesis,
very naturally yields the order of the field pressure variations
$O(x^{-2})$. 
Pressure variations were often overestimated at
$O(x^{-1})$\cite{Chang},\cite{TordellaAndBelan2003}.
This was  due to the use, in the inner expansion, of the assumption that the field can accommodate an inner pressure which is independent of the lateral coordinate, which however  varies at the leading orders along the $x$ coordinate.  However, at intermediate values of $y$ and for  fixed $x$, this assumption is responsible for an anomalous rise in the composite expansion, due to the central plateau that appears in the outer expansion.  The outer solution is in fact biased at finite values of $x$ to values greater than $1$ and forces the composite expansion to assume  inaccurate values  -- with respect to experimental results -- mostly in the region around $y/D \approx 2$ and outwards (at $y/D=20$ the longitudinal velocity is still appreciably different from $U$). For details, the reader may refer to Section IV and fig.6 in Tordella
and Belan 2003\cite{TordellaAndBelan2003}.


iv) Last, we would like to point out that we have used the Navier-Stokes equations in the whole field, without the addition of any further restrictive axiomatic position, such as the principle of exponential decay. This did not prevent our approximated solution from spontaneously showing the properties of rapid decay and irrotationality
at the first and second orders for the inner and the outer  flows, respectively. At the higher orders, which mainly influence the intermediate region, the decay becomes a fast algebraic decay.

For an unitary spanwise length, the defect of the volumetric flow rate $D$ is defined as
\begin{eqnarray}\label{volumetric_flow_rate}
    D(x) &=&  \int_{- \infty}^{+ \infty} (1 - u(x,y)) dy
\end{eqnarray}
\noindent and is approximated through $u_c = u_c(x,y)$, the composite solution for
the velocity field, as
\begin{eqnarray}\label{volumetric_flow_rate_approx}
    D(x) &\approx&  \int_{- \infty}^{+ \infty} (1 - u_c(x,y))
    dy.
\end{eqnarray}

Entrainment is the  quantity that takes into account the
variation of the volumetric flow rate  in the streamwise
direction, and is defined as

\begin{eqnarray}\label{entrainment_approx}
E(x) =
|\displaystyle{\frac{dD(x)}{dx}}|.
\end{eqnarray}

The sequence of the first four terms of the inner and outer
approximation for the streamwise velocity and the
transversal velocity is given in the following.



\noindent {\bf Zero order, n=0, }
\begin{eqnarray}
u_{i 0}(x,y) &=& c_0 \\
v_{i 0}(x,y) &=& 0 \\
u_{o 0}(x,y) &=& k_0\\
v_{o 0}(x,y) &=& 0
\end{eqnarray}

\noindent with $c_0 = 1, k_0 = 1$.


\noindent {\bf First order, n=1}
\begin{eqnarray}
u_{i 1}(x,y) &=& - A c_1 \textrm{e}^{- R y^2/(4x)} x^{-1/2} \\
v_{i 1}(x,y) &=& 0 \\
u_{o 1}(x,y) &=& 0 \\
v_{o 1}(x,y) &=& 0
\end{eqnarray}

\noindent with $c_1 = 1$, while the constant $A$ is related to the
drag coefficient ($A = \textstyle{\frac 1 4} (R / \pi)^{1/2}
c_D(R)$).


\noindent {\bf Second order, n=2}
\begin{eqnarray}
u_{i 2}(x,y) &=& - \frac{1}{2} A^2 \textrm{e}^{- R y^2/(4x)}
[\textrm{e}^{- R y^2/(4x)} + \frac{1}{2} \frac{y}{\sqrt x}
\sqrt{\pi R} \textrm{erf}(\frac{1}{2} \sqrt{\frac{R}{x}} y)]
x^{-1} \\
v_{i 2}(x,y) &=& - \frac{A}{2} \frac{y}{\sqrt x} \textrm{e}^{- R
y^2/(4x)}x^{-1} \\
u_{o 2}(x,y) &=& 0 \\
v_{o 2}(x,y) &=& 0
\end{eqnarray}

\noindent {\bf Third order,  n=3}
\begin{eqnarray}
u_{i 3}(x,y) &=& A^3 \textrm{e}^{- R y^2/(4x)} (2 - R \frac{y^2}{x})
[\frac{1}{2} c_3 - R F_3(x,y)] x^{-3/2} \nonumber \\
v_{i 3}(x,y) &=& - \frac{A^2}{2} \{- \frac{1}{2} \frac{y}{\sqrt x}
\textrm{e}^{- R y^2/(2x)}
  - \sqrt{\frac{\pi}{2R}} \textrm{erf}(\sqrt{\frac{R }{2 x}}y) + (\frac{1}{2}
  \sqrt{\frac{\pi}{R}} + \nonumber \\ &-& \frac{\sqrt{\pi R}}{4} \frac{y^2}{x}) \textrm{e}^{- R y^2/(4x)} \textrm{erf}(\frac{1}{2}
  \sqrt{\frac{R}{x}}y)\} x^{-3/2} \\
u_{o 3}(x,y) &=& Re(\frac{i}{3} k_{31} e^{(3i/2)arctan(s)} + k_{33}
\frac{s_+^{3/2}}{s^{3/2}} + \frac{1}{2} k_{32} s^{-3/2}s_+^{3/2}
\nonumber \\
&\times& \{ \frac{\sqrt{(1 + i s) s} (\frac{3}{4} -
\frac{i}{i+s})}{2(i+s)} + \frac{(-1)^{1/4}}{16 \sqrt 2} log
[\frac{(\frac{i-1}{\sqrt 2} + \sqrt s) (\frac{i-1}{\sqrt 2} -
(1-i) \sqrt{1 + i s} + \sqrt s)}{(\frac{1-i}{\sqrt 2} + \sqrt s)
(\frac{1-i}{\sqrt 2} + (1-i) \sqrt{1 + i s} + \sqrt s)}) \nonumber \\
v_{o 3}(x,y) &=& Re(e^{(3i/2)arctan(s)} [k_{31} + k_{32} s^+
\frac{s+i}{3(s-i)^2}]) \nonumber
\end{eqnarray}

\noindent where $c_3 = - 2.26605 + 0.15752 R - 0.00265 R^2 +
0.00001 R^3$, $F_3$ is the third order of the function
\begin{eqnarray}
F_n(x,y) = \frac{1}{\sqrt x} \int_0^y \frac{\textrm{e}^{R
\zeta^2/(4x)}}{\textrm{Hr}_{n-1}^2 (x,\zeta) G_n(x,\zeta)} d \zeta \\
G_n(x,y) = A^{-n} \frac{1}{\sqrt x} \int_0^y M_n(x,\zeta)
\textrm{Hr}_{n-1} (x,\zeta) d \zeta
\end{eqnarray}
\noindent where $ M_n(x,y)$ is the sum of the non homogeneous
terms of the general ordinary differential equation for the inner
solution coefficients ($\phi_n$), $n \ge 1$, obtained from the $x$
component of the Navier-Stokes equation\cite{BelanAndTordella2002}$^{,}$\cite{TordellaAndBelan2003}, and
$\textrm{Hr}_{n-1} (x,y) = \textrm{H}_{n-1}
(\displaystyle{\frac{1}{2} \sqrt{\frac{R}{x}} y})$, where
$\textrm{H}_{n}$ are Hermite polynomials. In the outer terms, the variables $r, s, s^{\pm}$  are defined as  $r = \sqrt{x^2+y^2}$, $s=y/x$, $s^{\pm} =
(1+s^2)^{\pm 1/2}$ and the relevant constants are  $k_{31} = \pm \frac{1}{2} A^2/2 \sqrt{\pi/(2R)}$,
$k_{32} = 3ik_{31}$, $k_{33} = 0$.




\section{Discussion of the results}


Before describing  the entrainment features we have observed, let us first discuss
the asymptotic behaviour of the inner expansion in the lateral far
field, since this aspect is important to determine the entrainment decay.
At finite values of $x$, the inner streamwise velocity decays to zero as a Gaussian
law for $n=1$ and as a power law of exponent $-2$ for $n=2$ and of
exponent $-3$ for $n \geq 3$. The cross-stream inner velocity  goes to zero for $n=0,1$ and to a
constant value for $n \geq 2$. This  allows $v$ to vanish as
$x^{-3/2}$ for $x \rightarrow \infty$. When $x \rightarrow \infty$
this approximation coincides with the Gaussian representation given by
the Oseen approximation. It can be concluded that, at Reynolds
numbers as low as the first critical value and where the
non-parallelism of the streamlines is not yet negligible, the
division of the field into two basic parts —- an inner vortical
boundary layer flow and an outer potential flow —- is
spontaneously shown up to the second order of accuracy ($n=1$). At
higher orders in the expansion, the vorticity is first
convected and then diffused in the outer field. This is  the
dynamical context in which the entrainment process takes place.

In figures 2 (a) and 3 (a), the longitudinal velocity profiles are
contrasted with the experimental data available for  steady flows
by Berrone\cite{berr-b},  Paranthoen et al.(1999)\cite{Paranthoen1999}, Nishioka \& Sato (1974)\cite{NishiokaSato} and Kovasznay (1948)\cite{kova}. The accuracy on the velocity distributions,
between the analytical data and the laboratory ones is lower than
5\%. This estimate was obtained by
contrasting the longitudinal velocity distribution $u$ with the
laboratory and numerical distributions, considered as the
reference distribution. To this end,  we computed the deviation $ \Delta_{ref} =
{\| u - u_{ref} \|_{0, x}} / {\|1-u_{ref}\|_{0,x} }$. At $R=34$, a
deviation $\simeq 4.5 \% \;$ was obtained for the laboratory results
by Kowasznay, where $x$ is the station at $20$ diameters from the
center of the cylinder, see fig.2. As for the data by Berrone,
we find a $\Delta_{ref}$ of about $1.7 \%$. At $x \sim 10$ we have
a comparison with Paranthoen et al. and Nishioka and Sato, that
yields a deviation $ \Delta_{ref}$ of about $2.5 \%$ and $1.5 \%$,
respectively, see fig.3.

The entrainment is closely linked to the lateral and far field asymptotic behaviour. Since a numerical  experiment cannot be over an unbounded domain of a flow, this approach is not suitable for the study of the field asymptotic behaviour and, as a consequence, entrainment (however, numerical simulations can yield very accurate representations of the near field, in particular of the standing eddy region). As far as entrainment is concerned, we then contrasted our analytical data with laboratory data. We tried to exploit all the results available in literature,
obtaining a comparison with Paranthoen et al. (1999)\cite {Paranthoen1999} and Kovasznay (1948)\cite {kova} because these authors present a sequence of velocity profiles (mainly in supercritical flow configurations) that extend into the intermediate wake. The comparison  was instead not feasible with the data by Nishioka and Sato (1974)\cite {NishiokaSato} since they mainly measured the near wake (standing eddy region). 

Figure 4 shows the volumetric flow rate defect
$D=D(x; R)$ and the entrainment $E=E(x; R)$  obtained from the composite expansion.
It can be observed that the volumetric defect flow rate slowly
decreases with the distance from the body (fig.4 a). This decrease
is faster at the beginning of the intermediate wake and at the
higher Reynolds values.
Considering  a fixed position $x$ (fig.4 c), the flow defect
decreases with the Reynolds  number. Fig.4 (a) includes data from
the laboratory experiments by Paranthoen et al. (1999, $R=53.3$)
and Kovasznay (1948, $R=56$),  both carried out at a slight
supercritical $R$ (unsteady regime). The difference between their results in not
small, but it should be recognized that the difficulties in
measuring at small values of the Reynolds
number are exceptionally high. By considering the
arithmetic mean between these two sets of data, an increase of more than
$50\%$  with regards to the values of the steady configuration, for $x < 20$, is
observed.

Parts (b, d) of fig.4 concern the entrainment, that is,
the spatial rate of change of the wake velocity defect. The
important points are: - the initial high variation at the
beginning of the intermediate part of the wake, which increases
with $R$, 
- the higher experimental mean value near $x=10$
($2.45 \; 10{-2}$ against $6.5 \; 10^{-3}$), - for all the $R$, the
exhaust of the entrainment at a distance of about $50$ body
lengths, - the collection of experimentally determined  values of
the critical $R$ number that has a median value of $46.6$: a fact
that relates the entrainment exhaust length - $EEL$ - to
$R_{cr}$ with a simple scaling, such as $EEL \; \sim \; {\displaystyle R_{cr}^{\, n}}$ with $n = 1$. In fig.4 (d), one can also
observe that at a constant distance $x$  from the body,  the entrainment
stops growing  beyond around  $R_{cr}$.

Though the connection between the entrainment length and the instability cannot be direct: - the first can be deduced as an integral property of the steady fully non linear version of the motion equations, - the second  from the  linear theory of stability, which is conceived to highlight the role of the perturbation characteristics and not of the integral properties of the basic flow,
these results could be {\it a fortiori} used to interpret the
bifurcation to the unsteady flow condition at $R_{cr}$  as a process  that allows  the wake to tune
the entrainment, and, possibly, to redistribute it on a larger wake
portion, according to the actual $R$ value.

It can be noticed that  the decay distance is of the same order of
magnitude as $R_{cr}$ and this shows that the scaling used in recent
stability analyses\cite{BelanAndTordella2006},\cite{TordellaAndScarsoglioAndBelan2006} to represent the slow time and
space wake evolution - $\tau=\varepsilon t$ and $\xi = \varepsilon
x$, where $\varepsilon = \displaystyle{\frac{1}{R} \sim \frac{1}{R_{cr}}}$ - is linked
to the exhaust of the entrainment process. In fact, one can say that the unit value of
the slow time and spatial scales is reached where the entrainment nearly
ends.

\section{Conclusions}

The entrainment is observed to be
intense in the intermediate wake downstream from the
separation region where the two-symmetric standing eddies are
situated. Here, the dependence on the Reynolds number is clear.
The entrainment grows six-fold when $R$ is increased from 20 to
100. The subsequent  downstream evolution presents a continuous
decrement of the entrainment. For all  the $R$ here considered, it has been observed that this decrease is
almost accomplished  at a distance from the body  of about $50$ diameters, which is a value that is close to the critical value $R_{cr}$ for the
onset of the first instability and the subsequent set up of the
unsteady regime (the median value in literature being $R_{cr} = 46.6$).  The establishment of
the unsteady regime could be interpreted as a way of overcoming the
limitation on the entrainment intensity and decay imposed
by the steady regime. The observed decay length  confirms the validity of the scaling that is often
adopted in wake stability studies carried out using the spatial
and temporal multitasking approach.


\vspace{3mm}
The authors would like to thank Marco Belan from the Politecnico di Milano for several helpful discussions.



\begin{figure}
\begin{minipage}[]{0.5\columnwidth}
   \includegraphics[width=\columnwidth]{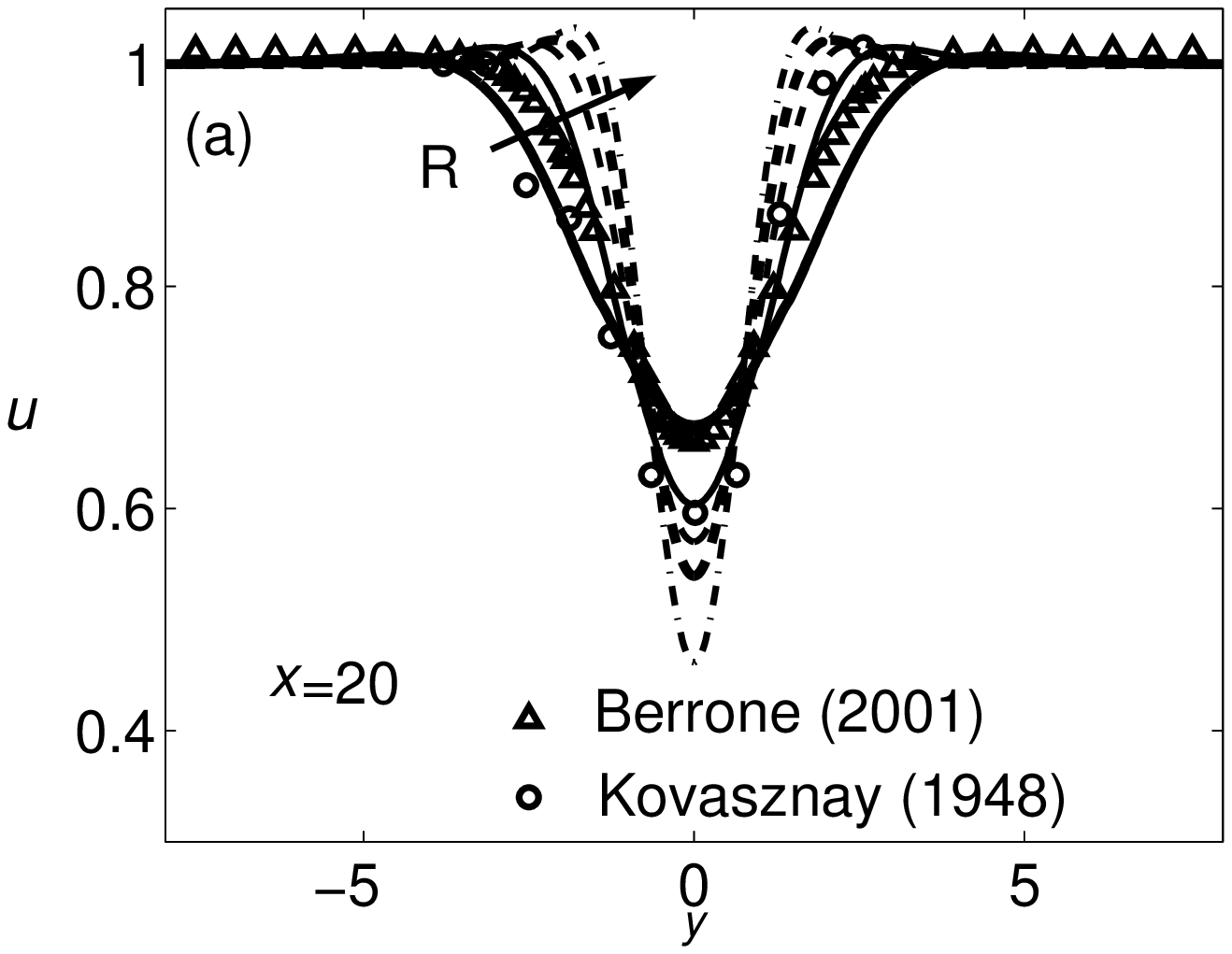}
    \label{JEM_wake_profile_base_flow_3}
\end{minipage}
\begin{minipage}[]{0.5\columnwidth}
   \includegraphics[width=\columnwidth]{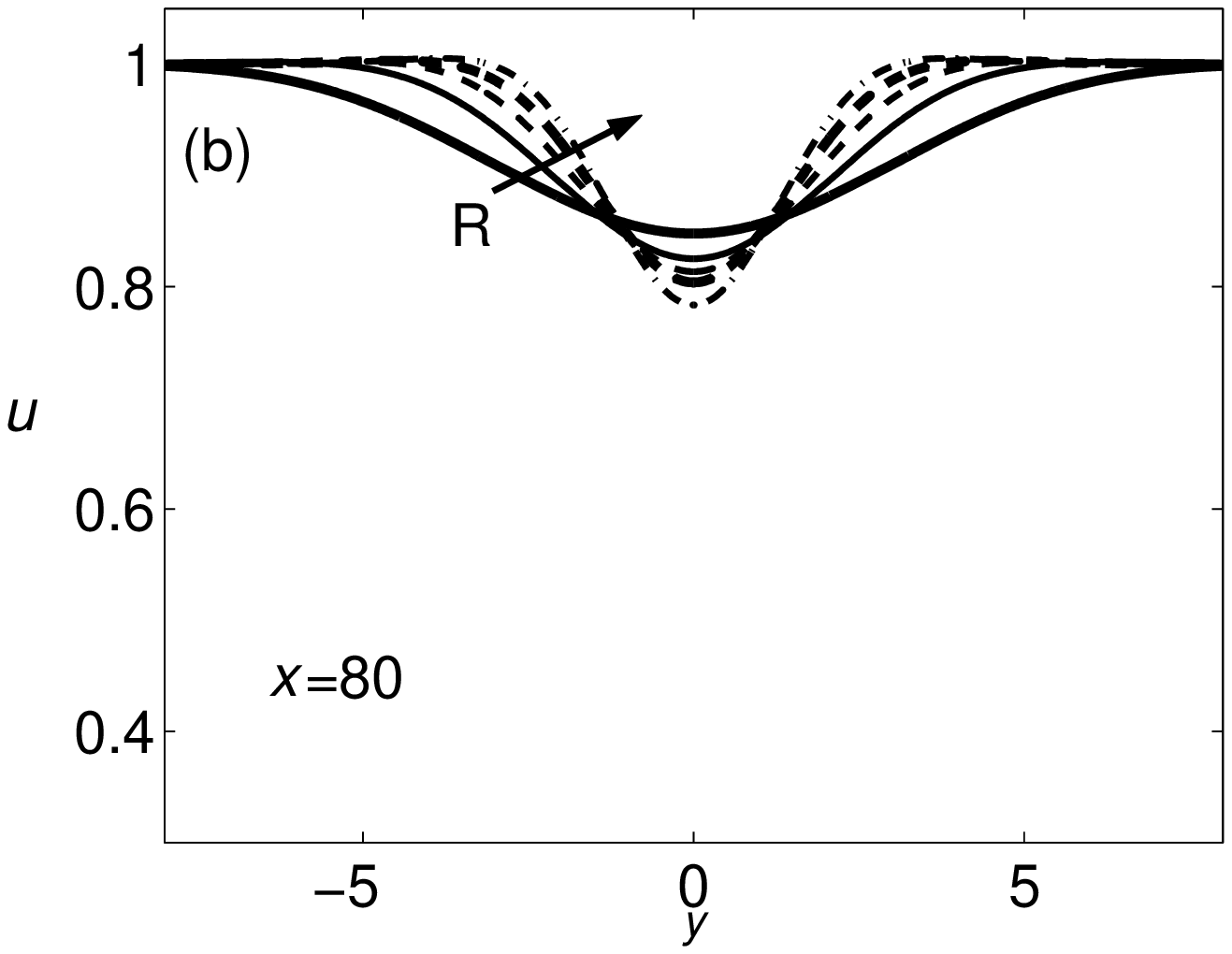}
    \label{JEM_wake_profile_base_flow_1}
\end{minipage}
\begin{minipage}[]{0.5\columnwidth}
   \includegraphics[width=\columnwidth]{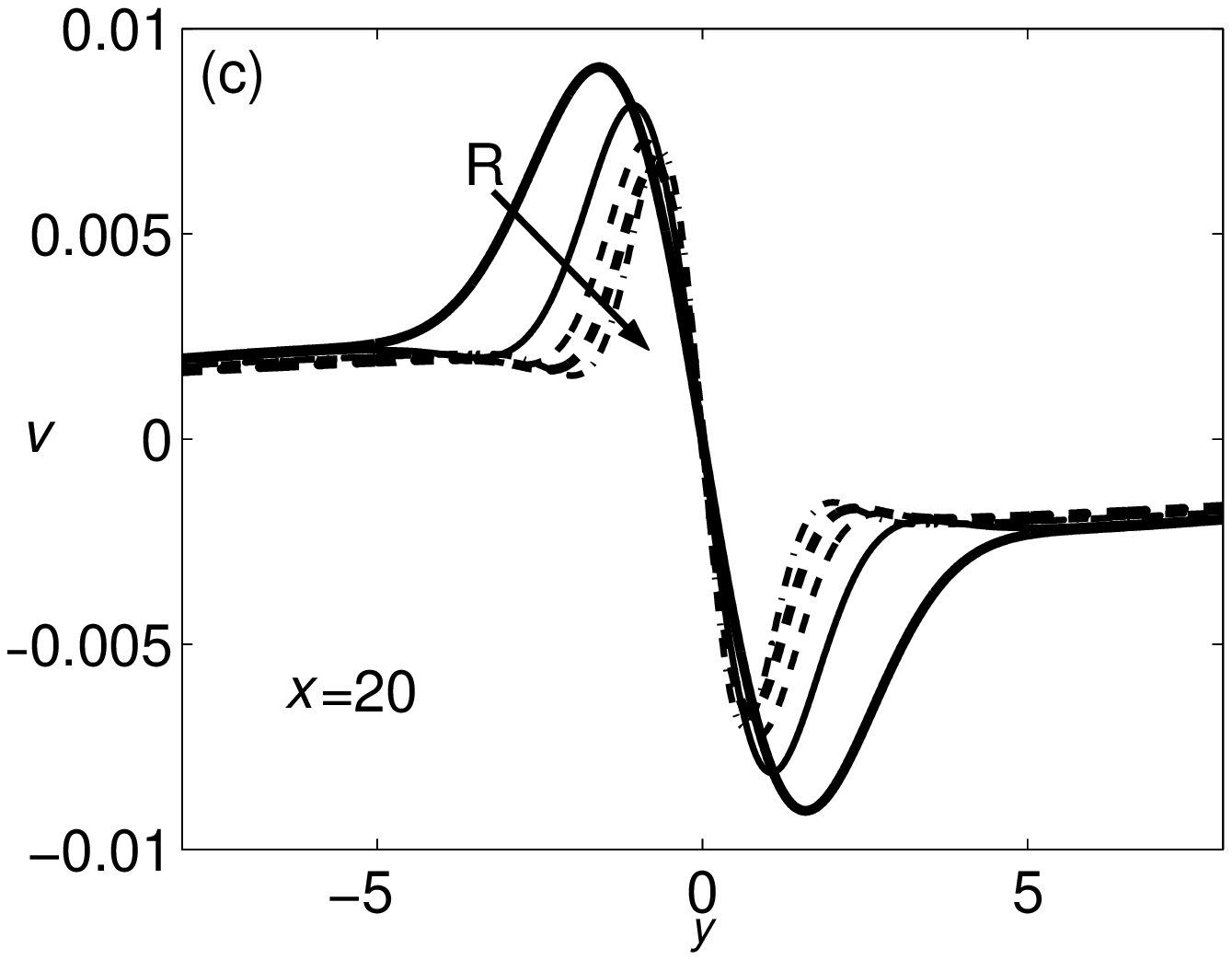}
    \label{JEM_wake_profile_base_flow_4}
\end{minipage}
\begin{minipage}[]{0.5\columnwidth}
   \includegraphics[width=\columnwidth]{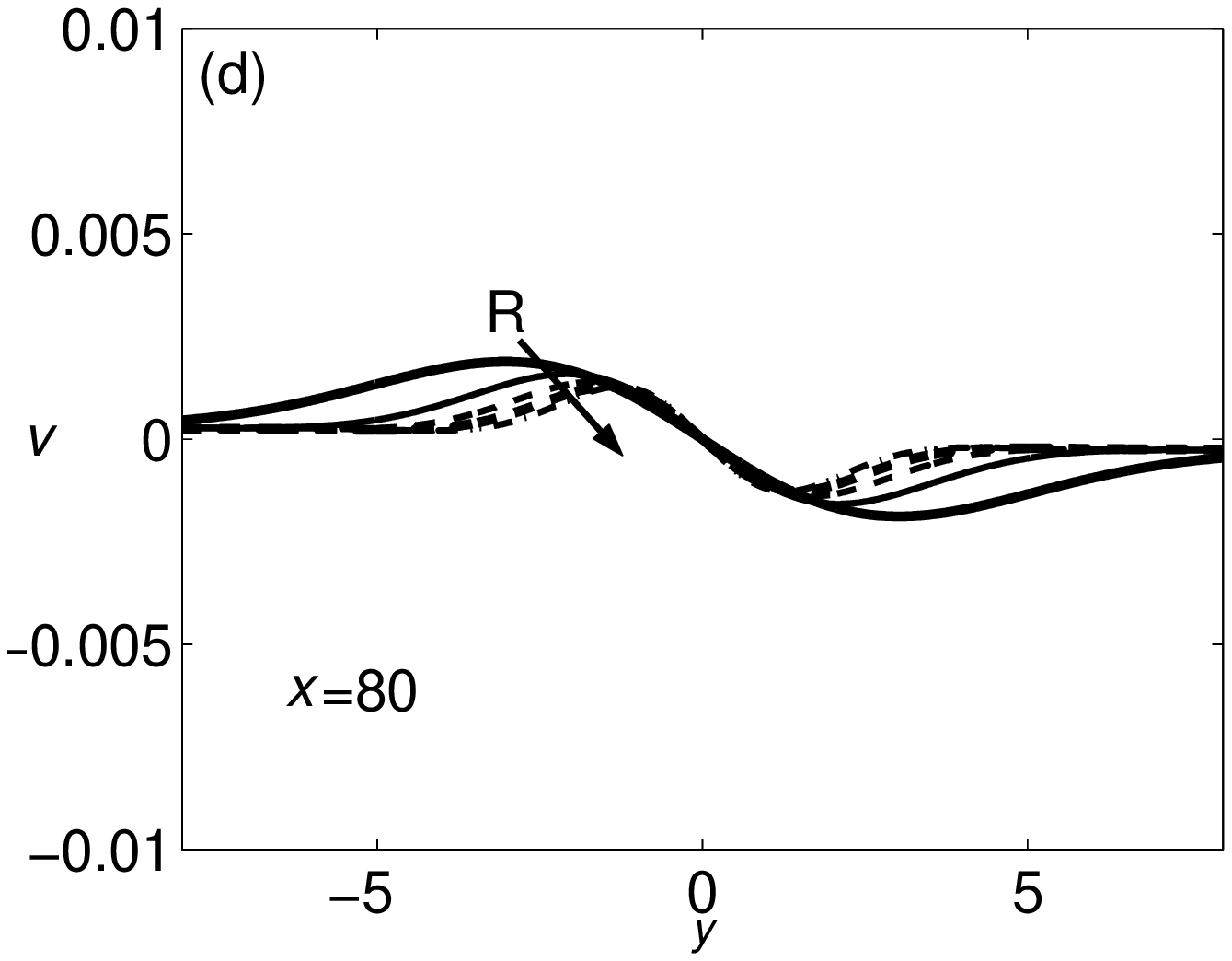}
    \label{JEM_wake_profile_base_flow_2}
\end{minipage}
\caption{- Velocity profiles at the downstream stations $x = 20$, $x
= 80$ and for $R = 20, 40, 60, 80$ and $100$. (a)-(b) Longitudinal
velocity $u$,  $x = 20$ and $x = 80$, (c)-(d) transversal velocity
$v$, $x = 20$ and $x = 80$. The comparison with the numerical results by
Berrone (2001)  (triangles, $R=34$, $x=20$) and the laboratory data by
Kovasznay (1948)  (circles, $R=34$, $x=20$) is shown in part (a).}
\label{fig_uv1}\end{figure}

\newpage

\begin{figure}
\begin{minipage}[]{0.5\columnwidth}
   \includegraphics[width=\columnwidth]{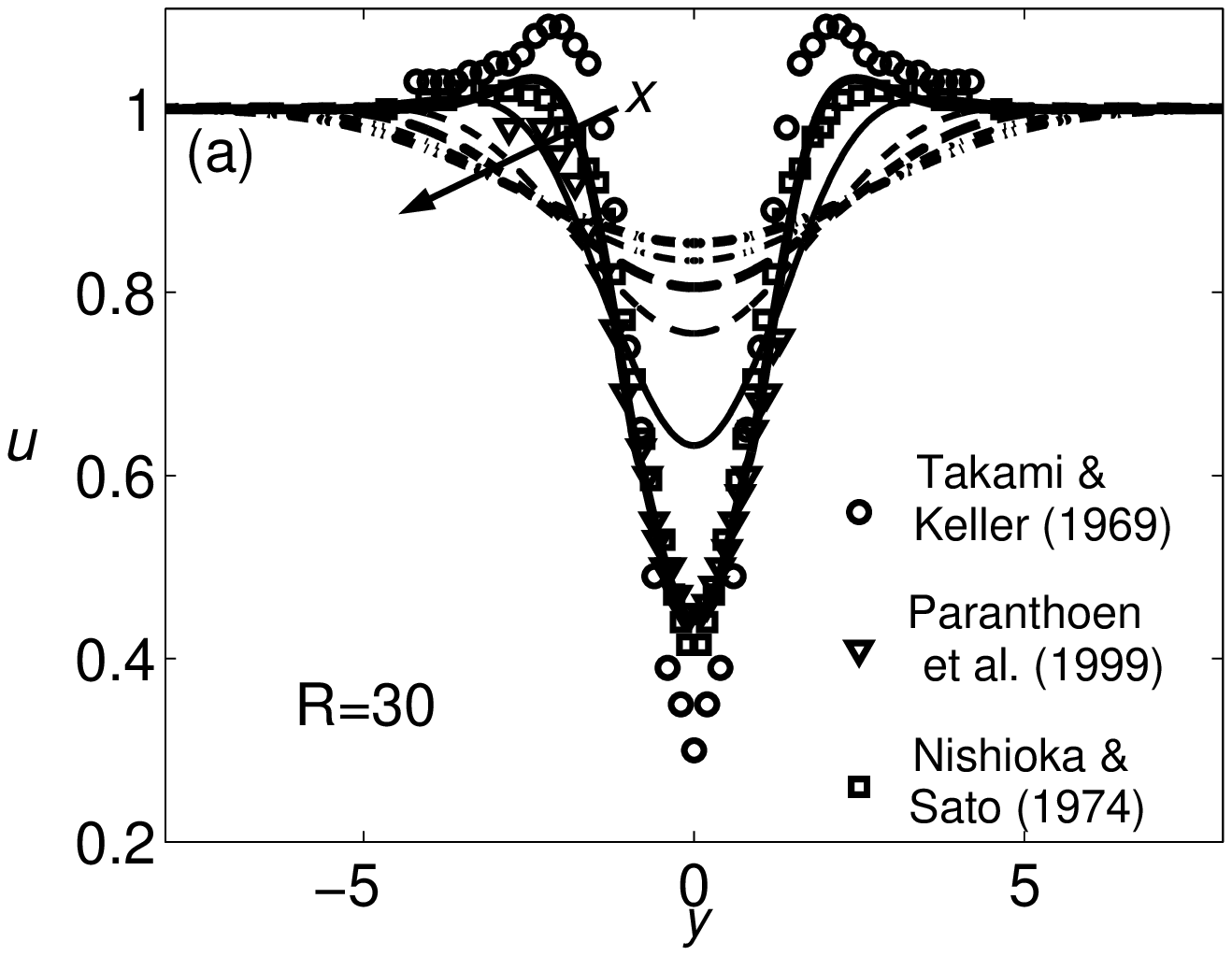}
    \label{JEM_u30}
\end{minipage}
\begin{minipage}[]{0.5\columnwidth}
   \includegraphics[width=\columnwidth]{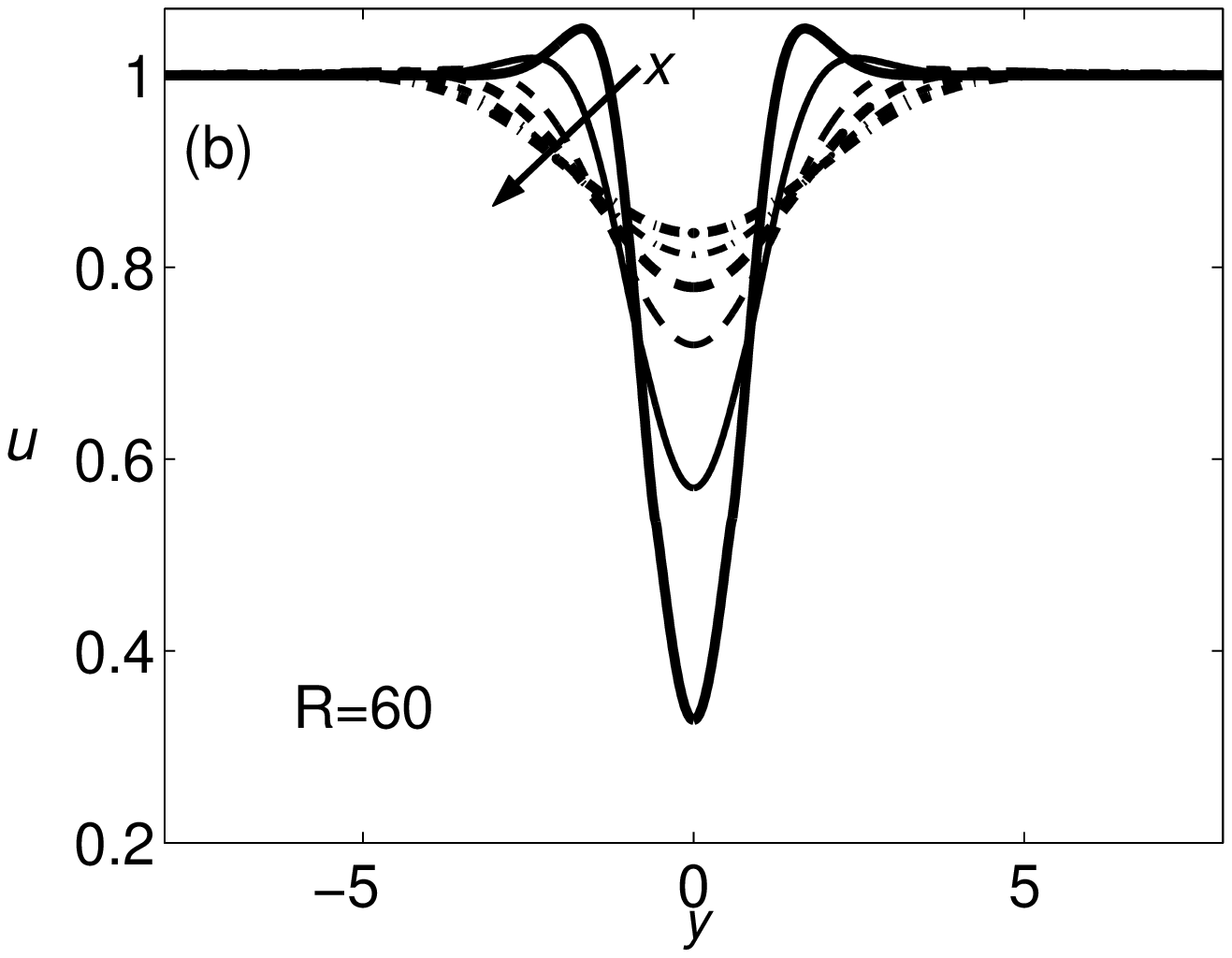}
    \label{JEM_u60}
\end{minipage}
\begin{minipage}[]{0.5\columnwidth}
   \includegraphics[width=\columnwidth]{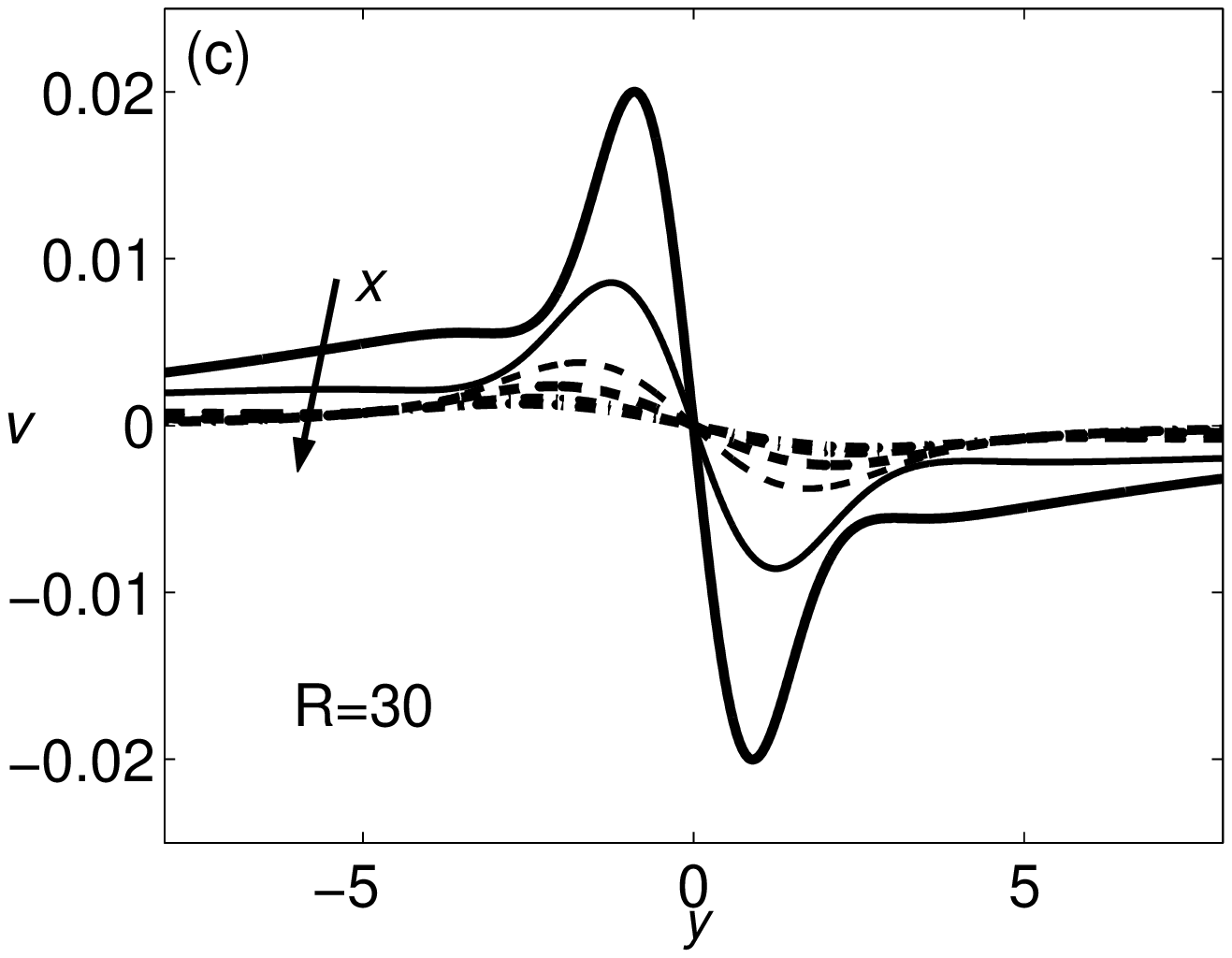}
    \label{JEM_v30}
\end{minipage}
\begin{minipage}[]{0.5\columnwidth}
   \includegraphics[width=\columnwidth]{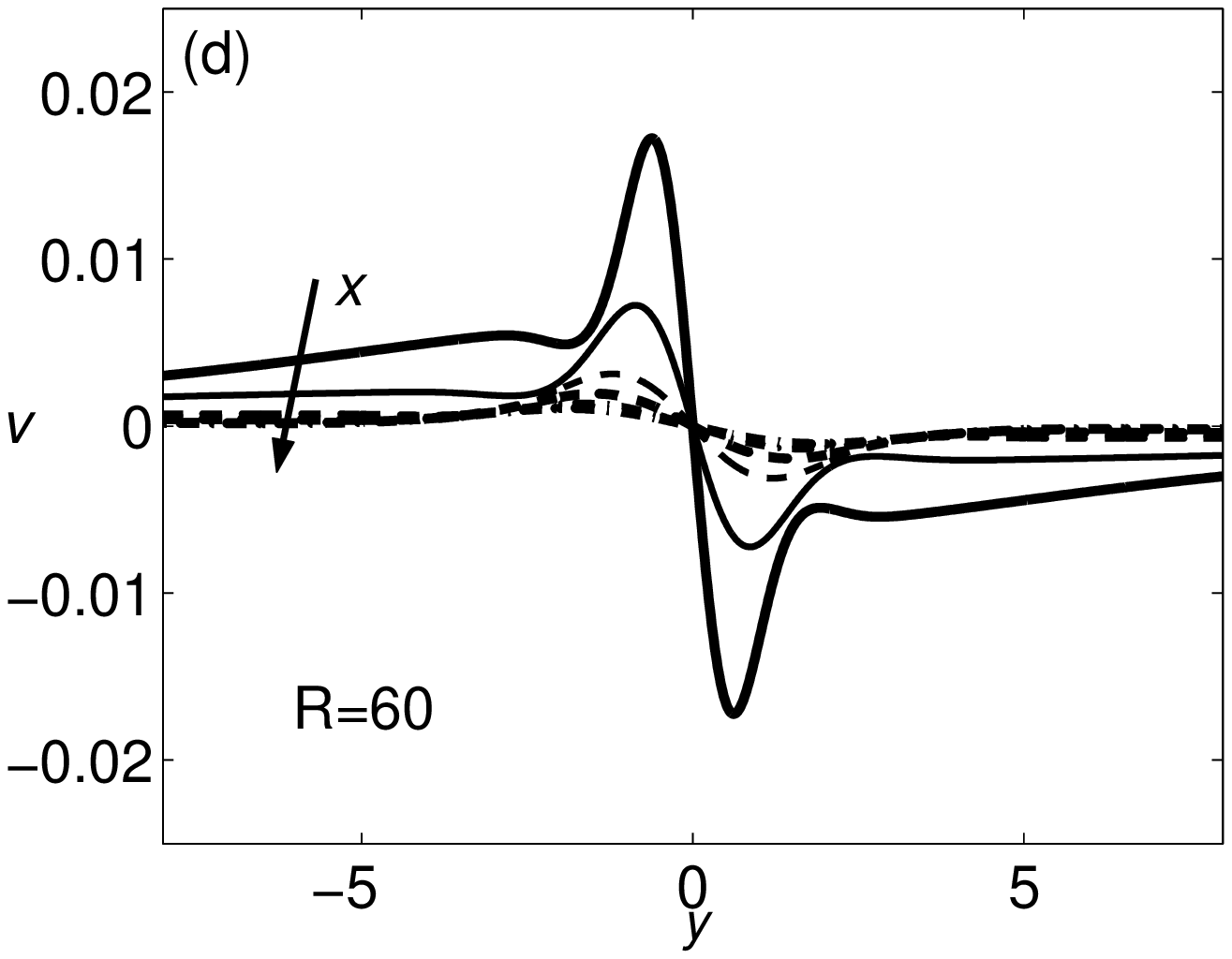}
    \label{JEM_v60}
\end{minipage}
\caption{- Velocity profiles for $R = 30, 60$ plotted at stations $x
= 10, 20, 40, 60, 80$ and $100$. (a)-(b) Longitudinal velocity $u$, $R
= 30$ and $R = 60$, (c)-(d) transversal velocity $v$, $R = 30$ and
$R = 60$. The comparison with the experimental data by Nishioka \& Sato (1974)
(squares, $R=40$, $x=7$) and Paranthoen et al.
(1999) (triangles, $R=34$, $x=10$) is shown in part (a).} \label{fig_uv}
\end{figure}

\newpage

\begin{figure}
\begin{minipage}[]{0.5\columnwidth}
   \includegraphics[width=\columnwidth]{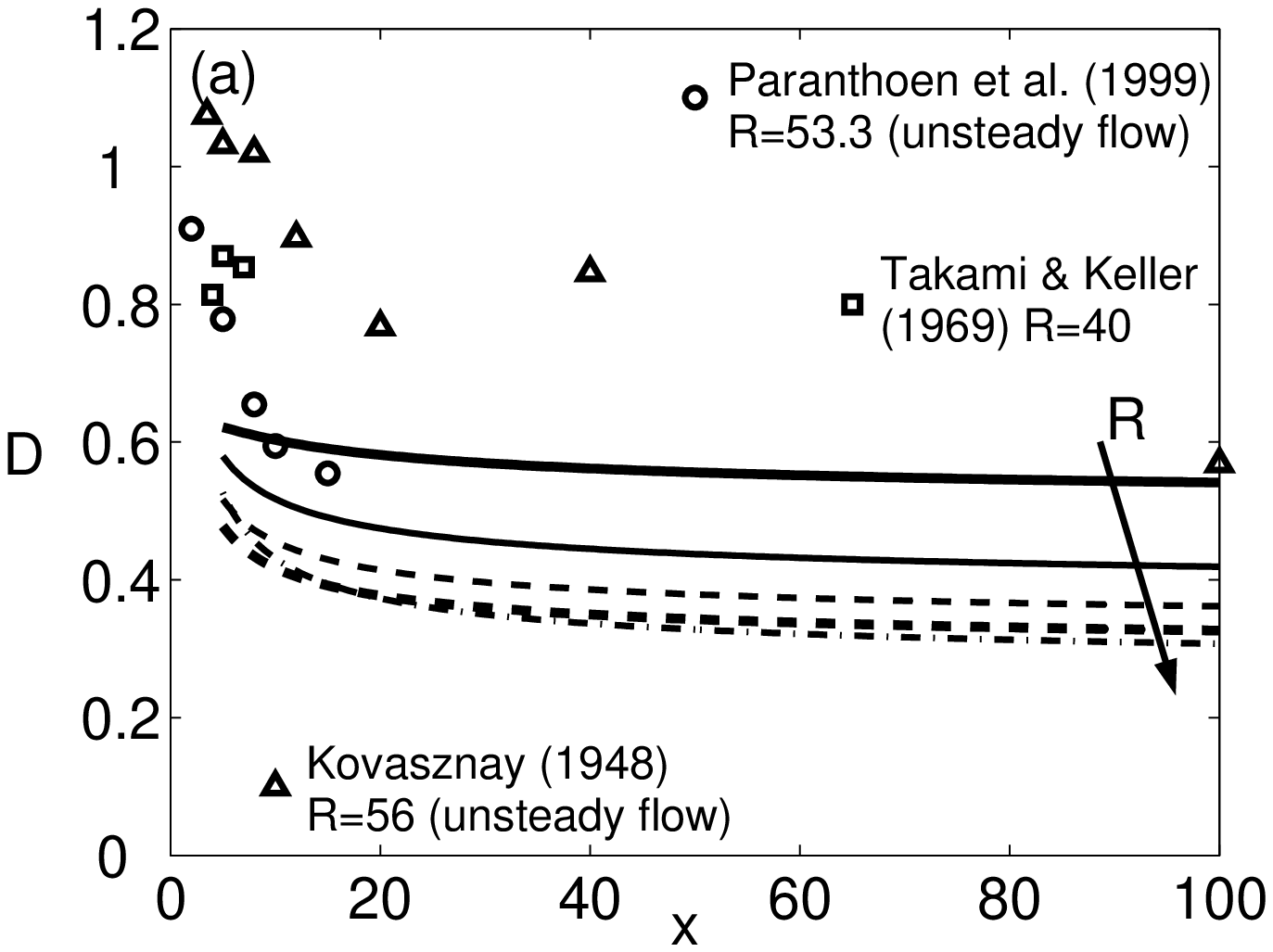}
    \label{D_x}
\end{minipage}
\begin{minipage}[]{0.5\columnwidth}
   \includegraphics[width=\columnwidth]{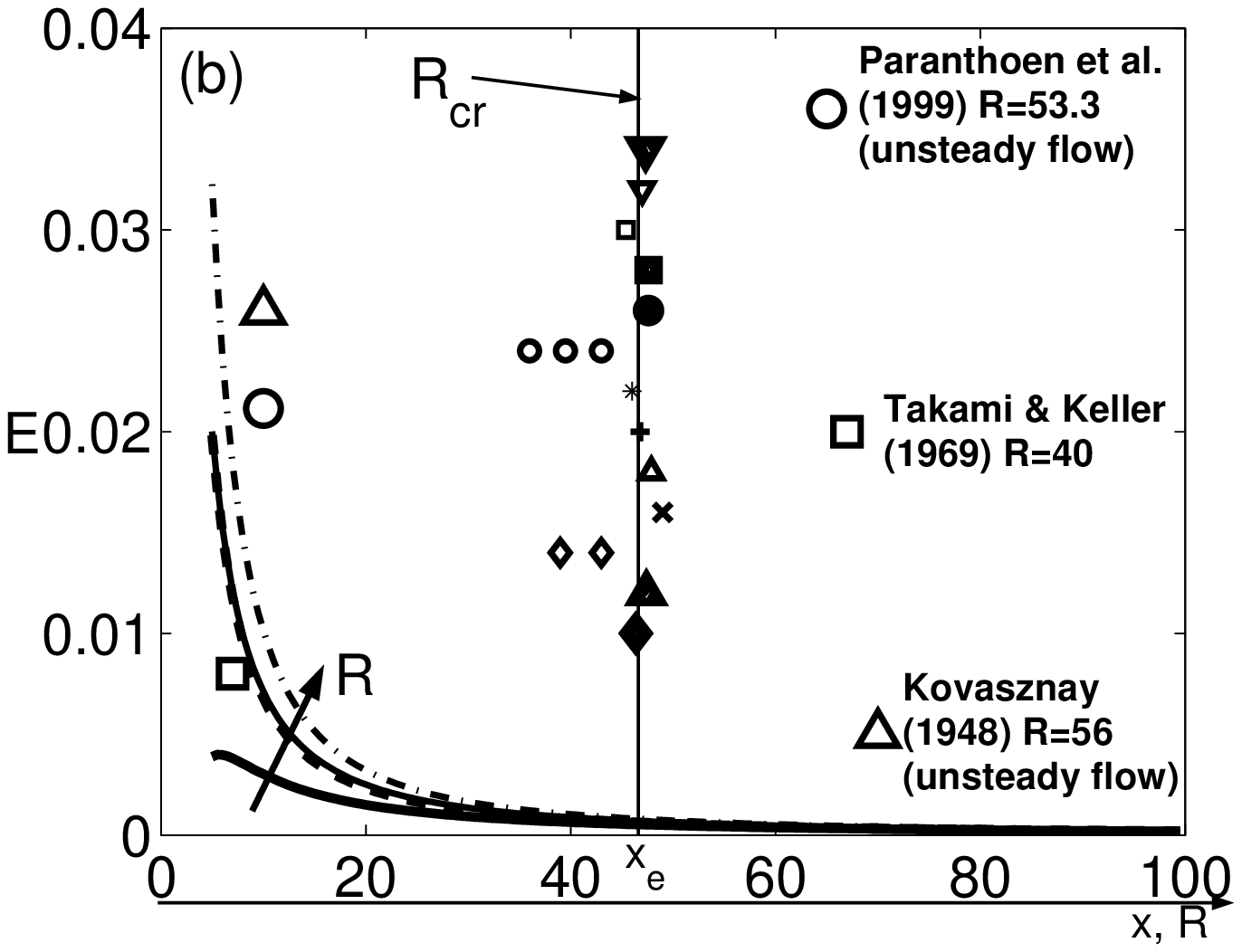}
    \label{E_x}
\end{minipage}
\vspace{-6mm}
\begin{minipage}[]{0.5\columnwidth}
   \includegraphics[width=\columnwidth]{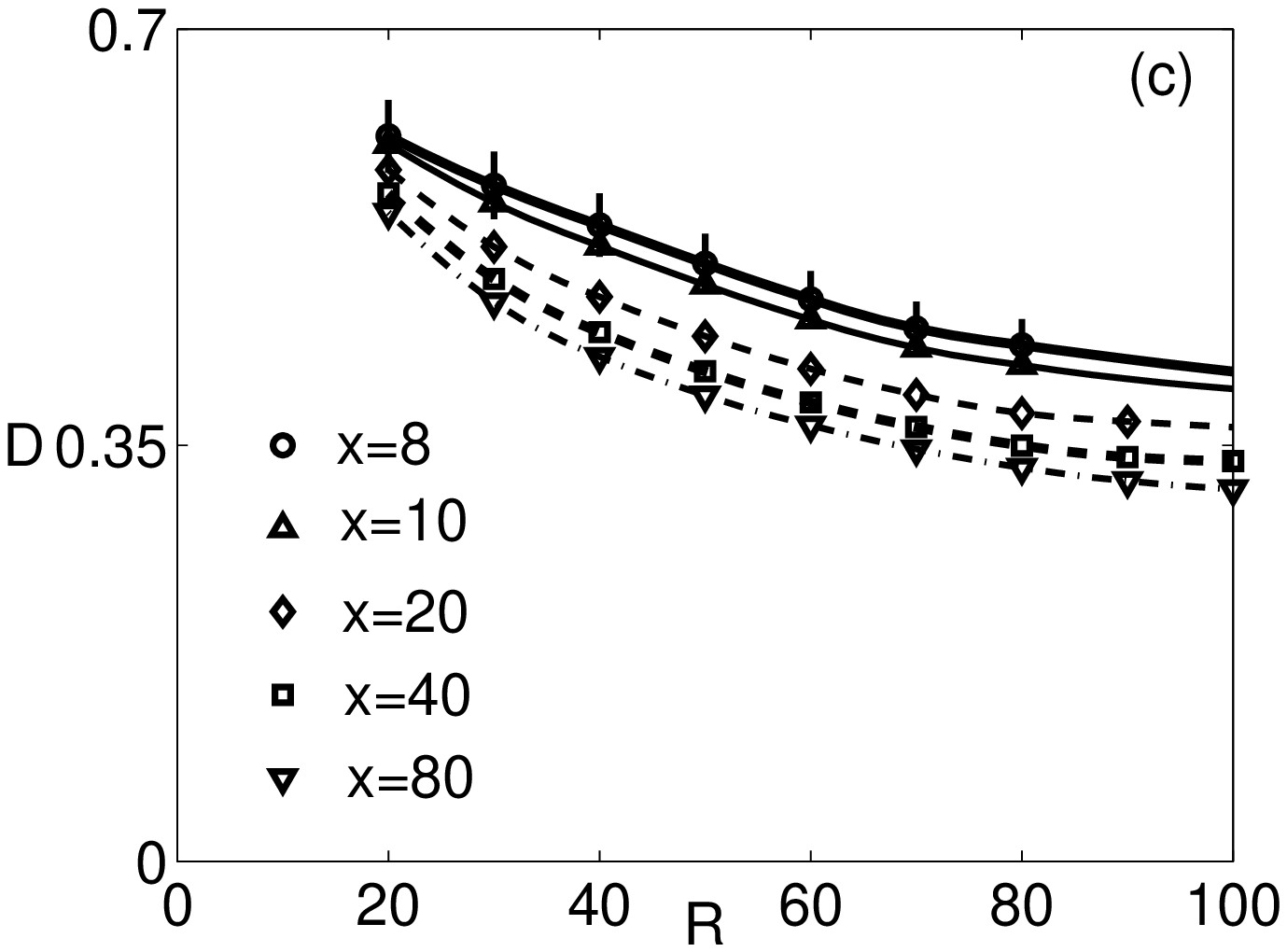}
    \label{D_Re}
\end{minipage}
\begin{minipage}[]{0.5\columnwidth}
   \includegraphics[width=\columnwidth]{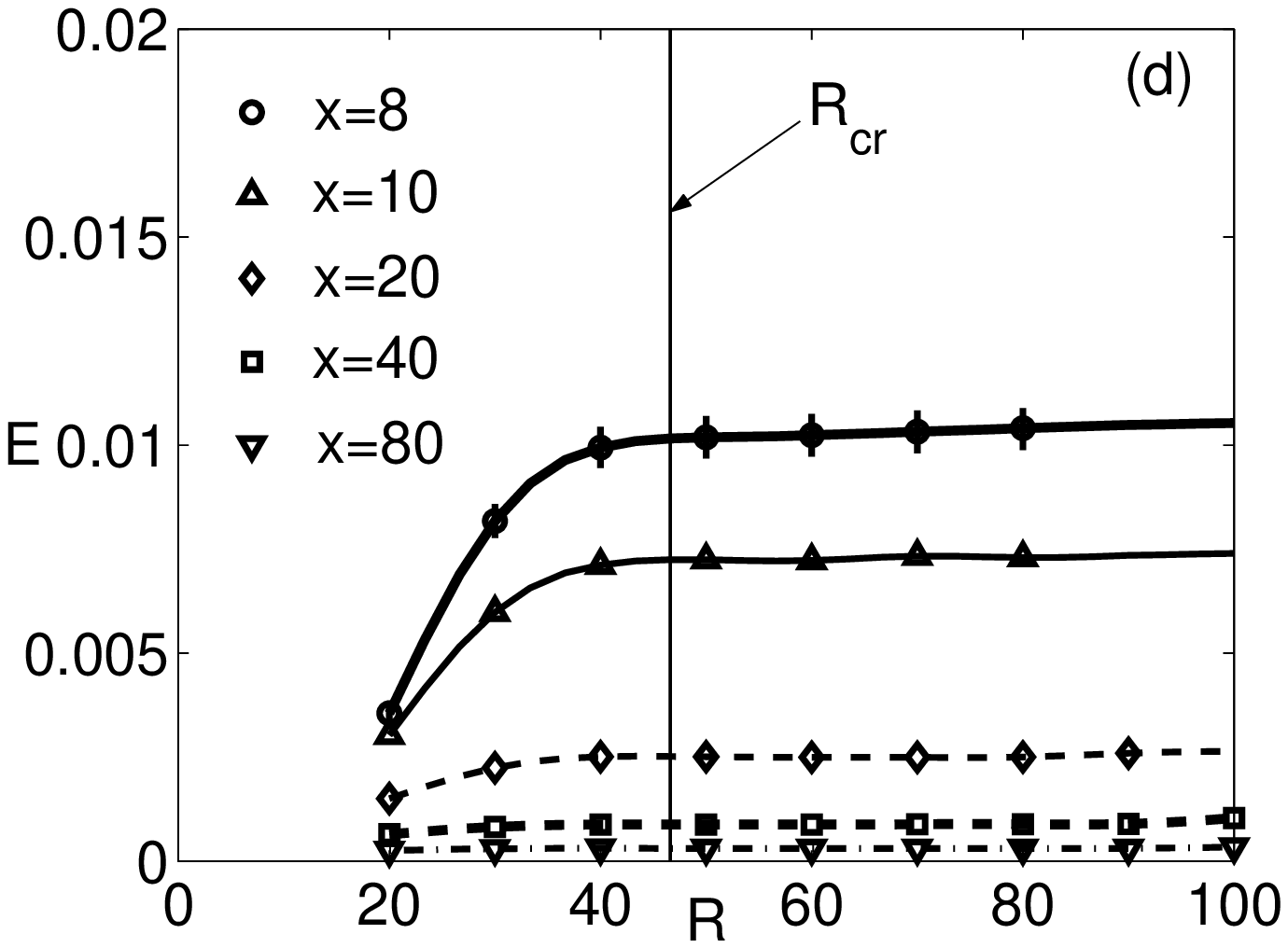}
    \label{E_Re}
\end{minipage}
\vspace{-6mm} \caption{\small - (a)-(b): Downstream distribution
of the volumetric flow rate defect $D$ and entrainment $E$ for $R
= 20, 40, 60, 80$ and $100$. (c)-(d): Volumetric flow rate defect
$D$ and entrainment $E$ as a function of the $R$ for different
stations ($x=8, 10, 20, 40, 80$). The values of the volumetric flow rate defect $D$ for the oscillating
(supercritical) wake, as inferred from experimental data by
Kovasznay (1948, $R=56$) and Paranthoen et al. (1999, $R=53.3$),
are also shown in part (a). The values of the critical Reynolds
number obtained from different numerical and experimental results
are placed at a distance from the body, $x_{e}$, equal to $
R_{cr}$, see  part (b). Position $x_{e}$ is observed to be the wake length where
the entrainment is almost extinguished $\forall R \in [20, 100]$, which leads to
the hypothesis that the steady wake
becomes unstable at a Reynolds number that is equal to  the normalized
distance where the entrainment almost ends and to the value beyond which the entrainment, at a constant distance from the body, stops growing (see part
(d)). The symbols represent data from:
Norberg 1994 ($\blacktriangle$), Zebib 1987 ($\lozenge$), Pier
2001 ($\times$), Williamson 1989 ($\vartriangle$), Leweke \&
Provansal 1995 ($+$), Strykowski \& Sreenivasan 1990 ($\ast$),
Coutanceau \& Bouard 1977 ($\circ$), Elsenlhor \& Eckelmann 1989
($\bullet$), Hammache \& Gharib 1989 ($\blacksquare$), Jackson
1987 ($\square$), Ding \& Kawahara 1999 ($\blacklozenge$),
Morzynski et al. 1999 ($\triangledown$), Kumar \& Mittal 2006
($\blacktriangledown$). The solid line in parts (b) and (d) indicates the median value
($R_{cr} \approx 46.6$) of these data. } \label{DE}
\end{figure}

\end{document}